\documentclass{emulateapj}
\usepackage{times}
\newcommand\bb[1]{\mbox{\boldmath{$#1$}}}
\newcommand\del{\bb{\nabla}} 
\newcommand\bcdot{\bb{\cdot}}

\begin{document}

\shorttitle{\textsc SATURATION OF MRI VIA PARASITIC MODES}
\shortauthors{\textsc Pessah \& Goodman}

\title{On the Saturation of the Magnetorotational Instability via Parasitic Modes}
\author{Martin E. Pessah}
\affil{Institute for Advanced Study, Princeton, NJ, 08540}
\and 
\author{Jeremy Goodman}
\affil{Princeton University Observatory, Princeton, NJ 08544}
\email{mpessah@ias.edu, jeremy@astro.princeton.edu}

\begin{abstract}
  We investigate the stability of incompressible, exact, non-ideal
  magnetorotational (MRI) modes against parasitic instabilities.  Both
  Kelvin-Helmholtz and tearing-mode parasitic instabilities may occur
  in the dissipative regimes accessible to current numerical
  simulations.  We suppose that a primary MRI mode saturates at an
  amplitude such that its fastest parasite has a growth rate
  comparable to its own.  The predicted alpha parameter then depends
  critically on whether the fastest primary and parasitic modes fit
  within the computational domain and whether non-axisymmetric
  parasitic modes are allowed.  Hence even simulations that resolve
  viscous and resistive scales may not saturate properly unless the
  numerical domain is large enough to allow the free evolution of both
  MRI and parasitic modes.  To minimally satisfy these requirements in
  simulations with vertical background fields, the vertical extent of
  the domain should accommodate the fastest growing MRI mode while the
  radial and azimuthal extents must be twice as large. The
  fastest parasites have horizontal wavelengths roughly twice as long
  as the vertical wavelength of the primary.
\end{abstract}


\section{Introduction}
\label{sec:introduction}

Understanding the processes that halt the exponential growth of the
magnetorotational instability (MRI; \citealt{Velikhov59,
  Chandrasekhar60, BH91,BH98}) and set the rate of angular momentum
transport in the turbulent regime has been an outstanding problem in
accretion physics for almost two decades.  The net magnetic flux
\citep{HGB95, Sanoetal04, PCP07}, the geometry of the domain
\citep{HGB95, Bodoetal08}, the resolution \citep{PCP07, FP07}, and the
microphysical dissipation coefficients \citep{FSH00, SI01, FPLH07,
  LL07, MS08}, all influence the non-linear saturation of the MRI in
simulations.

Saturation of the MRI may be related to secondary (parasitic)
instabilities \citep[][hereafter GX94]{GX94} that feed upon the free
energy afforded by the MRI (see, e.g.,
\citealt{KJ05,UMR07,TD08,Vishniac09} for alternative ideas). These
instabilities are often invoked to explain some of the behavior
observed in numerical simulations.  The absence of explicit
dissipation from most numerical studies and from GX94's analysis,
however, impedes quantitative interpretations.

Both primary and secondary instabilities may be subject to non-ideal
effects.  The growth rates and wavenumbers of the fastest growing MRI
primaries, and the relative orientations and magnitudes of their
velocity and magnetic field perturbations, are sensitive to
dissipation coefficients (see, e.g., \citealt{LL07, LB07, PC08}).
Viscosity may slow the growth of the Kelvin-Helmholtz secondaries
identified by GX94, and non-zero resistivity may enable resistive
instabilities such as tearing modes (see, e.g., \citealt{BS03}).

Here we summarize a parametric study of parasitic instabilities in
dissipative regimes accessible to current numerical simulations. We
adopt the incompressible limit, which is relevant for initial fields
so weak that saturation occurs with sub-equipartition fields. The
fastest growing, non-ideal parasitic modes are related to
Kelvin-Helmholtz and tearing-mode instabilities. They are
non-axisymmetric and have horizontal wavelengths roughly a factor of
2 larger than the vertical wavelength of the primary MRI mode. Our
findings suggest that current simulation domains may bias the
saturation of the non-ideal MRI by excluding the fastest MRI and
parasitic modes (see also \citealt{HGB95, Sano07, Bodoetal08}).

\begin{figure*}[t]
  \includegraphics[width=2\columnwidth,trim=0 0 0 0]{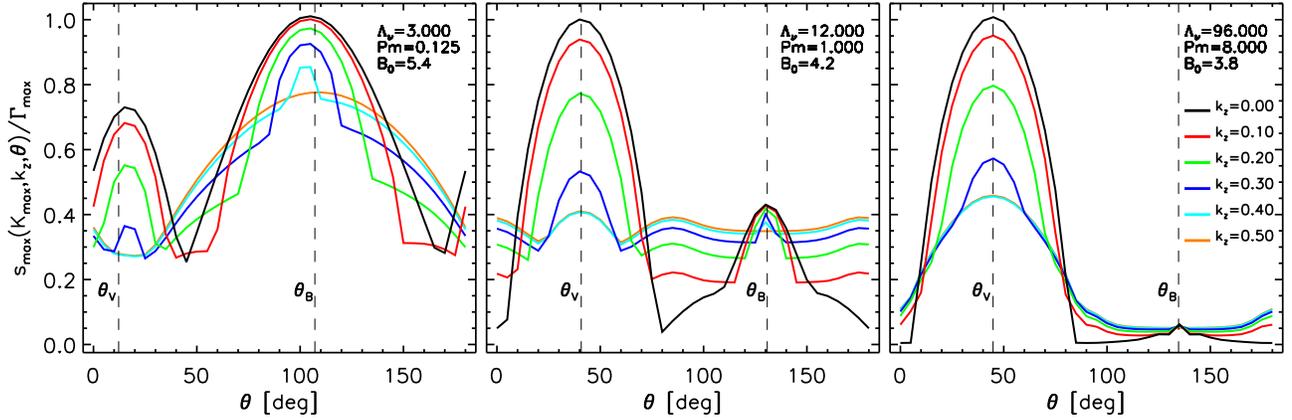}
  \caption{Normalized growth rates of the fastest growing parasitic
    modes vs. the orientation of the horizontal wavevector
    $\bb{k}_{\rm h}$ with respect to the radial ($\theta=0$)
    direction.  In each panel, the MRI magnetic field $B_0=B_0^{\rm
      sat}(\nu,\eta)$ is such that the fastest parasitic growth rate,
    maximized over $k_{\rm h}$, $\theta$ and $k_z$, matches the growth
    rate of the fastest primary MRI mode, $\Gamma_{\rm
      max}(\nu,\eta)$, for the indicated ``viscous Elsasser'' and
    magnetic Prandtl numbers, $\Lambda_\nu$ and ${\rm Pm}$.  The
    angles $\theta_{\rm V}$ and $\theta_{\rm B}$ mark the directions
    of the horizontal velocity and magnetic fields of the fastest MRI
    mode.  The fastest parasites lie at $\theta\approx\theta_{\rm V}$
    or $\theta\approx\theta_{\rm B}$ and are associated with
    Kelvin-Helmholtz and tearing mode instabilities,
    respectively. Tearing modes gain prominence at lower ${\rm Pm}$.}
  \label{fig:max_growths}
\end{figure*}

\section{Viscous, Resistive Primary MRI Modes}
\label{sec:assumptions}

Consider an incompressible, Keplerian
background with constant viscosity $\nu$ and resistivity $\eta$ and
threaded by a vertical magnetic field, $\bar{B}_z$, against
perturbations parallel to the background magnetic field.  \citet{PC08}
showed that in the shearing box, exact unstable solutions exist of the form
\begin{eqnarray}
\label{eq:V_exact}
\bb{v}&=&- q \Omega_0 (r-r_0) \hat{\bb{\phi}} +  \bb{V}_0 \sin(Kz) \, e^{\Gamma t} \,,\\ 
\label{eq:B_exact}
\bb{B}&=& \bar{B}_z \hat{\bb{z}} +  \bb{B}_0 \cos(Kz) \, e^{\Gamma t} \,,
\end{eqnarray}
where $q\equiv-d\ln\Omega/d\ln r$ and $\Omega_0$ is the local
Keplerian frequency.  For a given wavenumber $K$, the growth rate
$\Gamma$ satisfies the dispersion relation
\begin{equation}
\label{eq:dispersion_relation_nu_eta}
(K^2\bar{v}_{{\rm A}z}^2 + \Gamma_\nu \Gamma_\eta)^2 
+ \kappa^2 (K^2\bar{v}_{{\rm A}z}^2 + \Gamma_\eta^2) 
- 4 K^2\bar{v}_{{\rm A}z}^2\Omega_0^2 = 0 \,,
\end{equation}
where $\Gamma_\nu \equiv \Gamma + \nu K^2$, $\Gamma_\eta \equiv \Gamma
+ \eta K^2$, $\kappa\equiv\sqrt{2(2-q)}\Omega_0$ is the epicyclic
frequency, $\bar{v}_{{\rm A}z}\equiv\bar{B}_z/\sqrt{4\pi\rho}$ is the
Alfv\'en speed, and $\rho$ is the density. The relative strength
$V_0/B_0$ of the MRI velocity and magnetic fields, $\bb{V}_0 =
V_0\,(\cos\theta_{{\rm V}}, \sin\theta_{{\rm V}}, 0)$ and $\bb{B}_0 =
B_0\,(\cos\theta_{{\rm B}}, \sin\theta_{{\rm B}}, 0)$, and their
directions $\theta_{{\rm V}}$ and $\theta_{{\rm B}}$, are known
functions of $(\nu,\eta,K)$. The growth rate $\Gamma$ has a unique
maximum, $\Gamma_{\rm max}(\nu,\eta)$, at $K=K_{\rm max}(\nu,\eta)$.

Numerical simulations impose periodicity lengths on the shearing
box. We avoid finite-volume effects in our analysis by adopting length
and time scales based on intensive parameters: $L_0 \equiv
\bar{v}_{{\rm A}z}/\Omega_0$ and $T_0 \equiv 1/\Omega_0$. Viscosity
and resistivity introduce two new scales, which we subsume into the
dimensionless quantities $\Lambda_\nu \equiv \bar{v}_{{\rm
    A}z}^2/\nu\Omega_0$ and $\Lambda_\eta \equiv \bar{v}_{{\rm
    A}z}^2/\eta\Omega_0$, whose ratio is the magnetic Prandtl number,
${\rm Pm} \equiv \Lambda_\eta/\Lambda_\nu = \nu/\eta$. The quantity
$\Lambda_\eta$ is known as the Elsasser number, while its viscous
counterpart $\Lambda_\nu$ is related to the Reynolds number (see
below).  Throughout the rest of the Letter, unless otherwise mentioned,
we use the scales $L_0$ and $T_0$ to define dimensionless
variables. In these units, magnetic field strengths are defined
relative to the background field $\bar{B}_z$, while $\Lambda_\nu \to
\nu^{-1}$ and $\Lambda_\eta \to \eta^{-1}$.

\section{Viscous, Resistive Parasitic Modes}
\label{sec:parasitic instabilities}

To facilitate our analysis of the stability of the non-ideal MRI modes
against parasitic instabilities, we invoke some approximations
similar to those adopted by GX94 for ideal magnetohydrodynamics (MHD).
We assume that the amplitudes of the primary modes are large enough
that we can neglect the influence of the weak vertical background
field, the Coriolis force, and the background shear flow on the
dynamics of the secondary modes. Furthermore, the energy source for
secondary instabilities should increase with (some power of) the
amplitudes of the primary modes. Therefore, the secondary growth rates
should eventually outstrip the primary ones. We thus assume, following
GX94, that we can neglect the temporal variation of the sinusoidal
velocity and magnetic fields of the primary modes.

\begin{figure*}[t]
  \includegraphics[width=\columnwidth,trim=0 0 0 0]{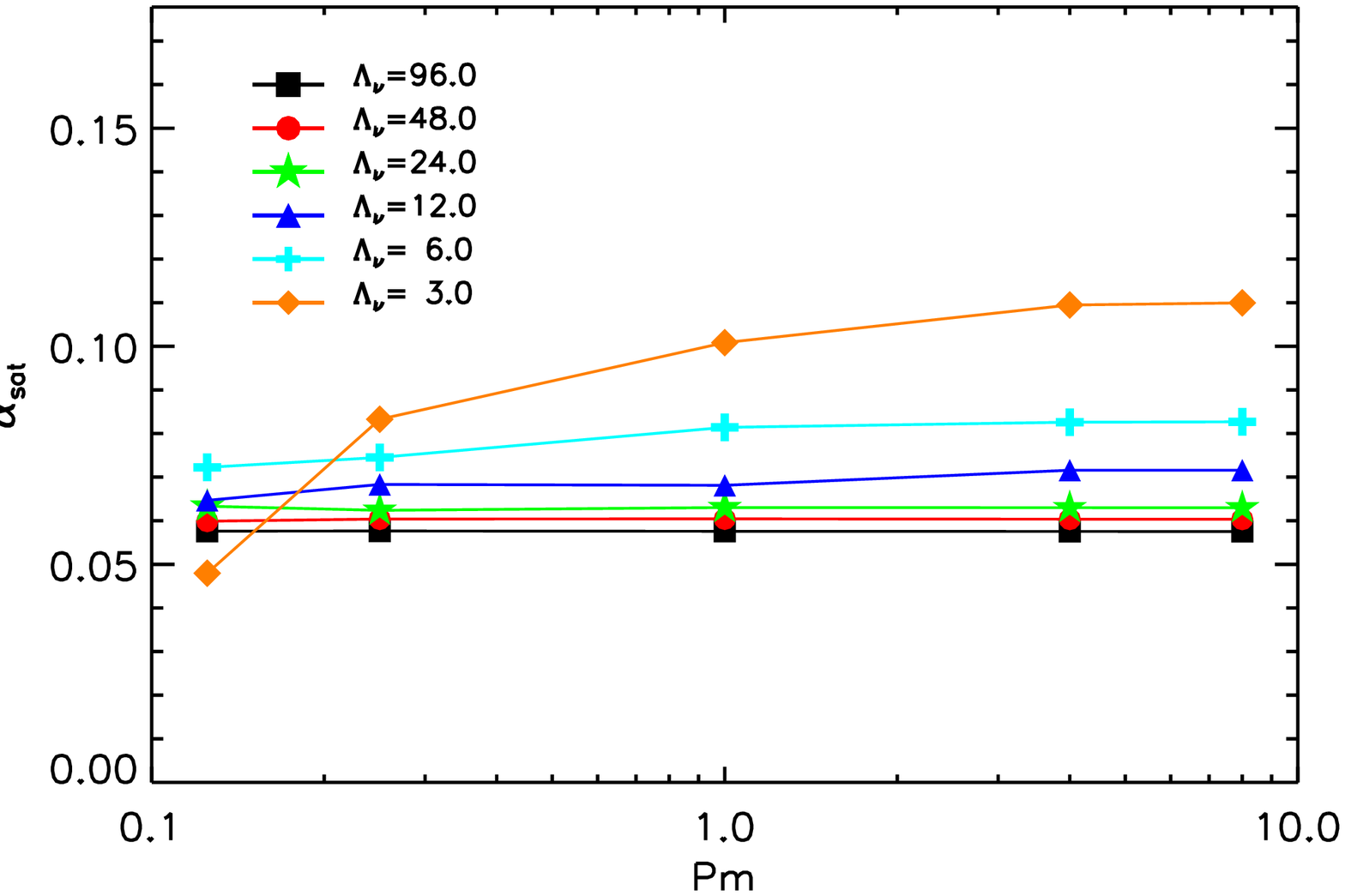}
  \includegraphics[width=\columnwidth,trim=0 0 0 0]{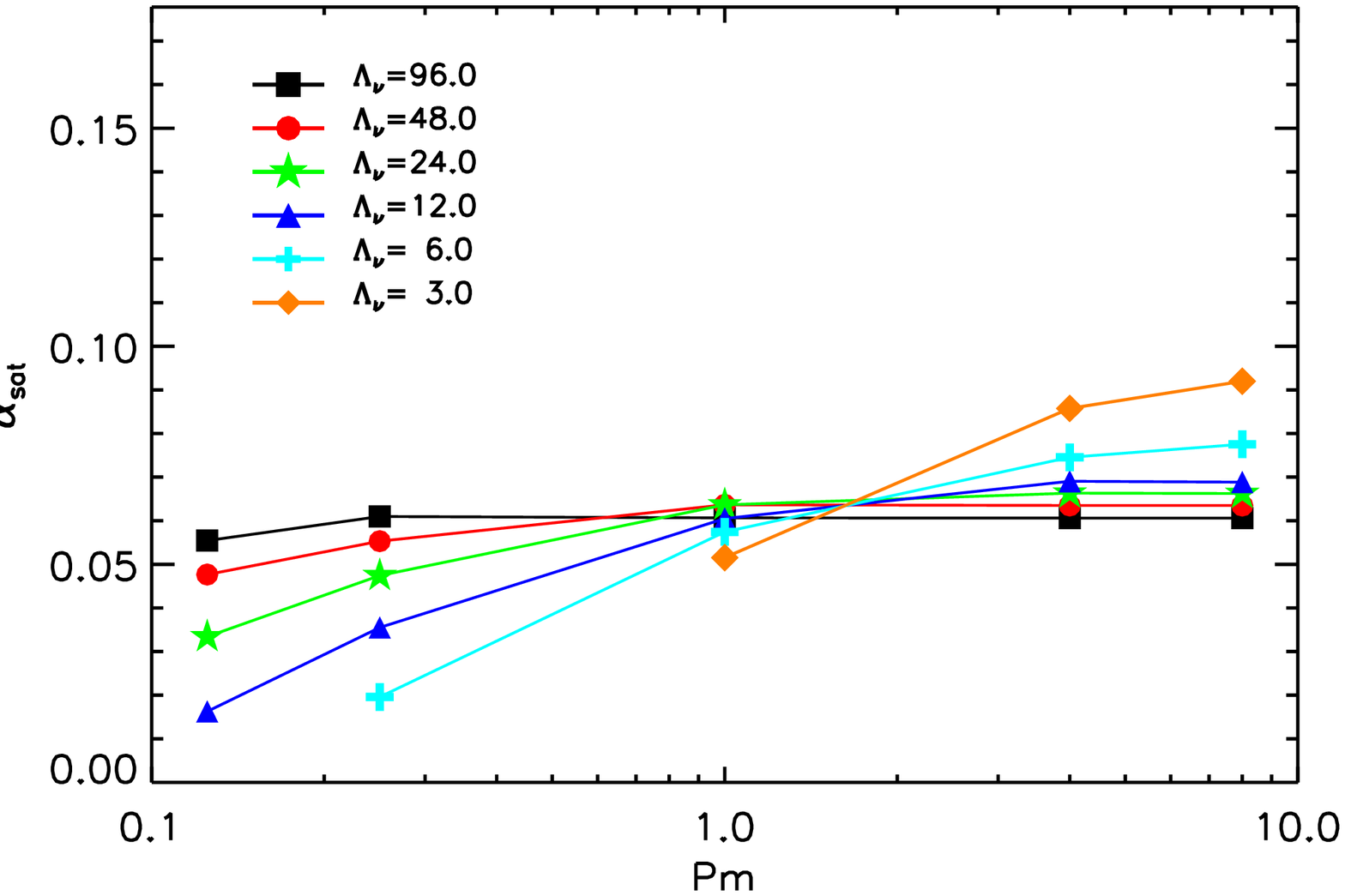}
  \caption{Predicted dimensionless stress at saturation, $\alpha_{\rm
      sat}$, as a function of $\Lambda_\nu$ and ${\rm Pm}$, if
    saturation occurs when the fastest parasitic and primary MRI
    growth rates match.  The indicated values for $\Lambda_\nu$
    correspond to the Reynolds numbers studied by LL07, ${\rm Re} =
    \{200, 400, 800, 1600, 3200, 6400\}$. In the left panel the
    fastest MRI and parasitic modes are allowed to evolve
    unimpeded. In the right panel the primary and secondary modes
    considered are the fastest MRI and parasitic modes that can fit in
    a domain with $(L_r,L_{\phi},L_z) = (1,4,1) \times
    2\sqrt{\beta}/3$ with $\beta=100$, as considered in LL07.}
  \label{fig:alpha_sat}
\end{figure*}

Under these assumptions, the equations governing the dynamics of the
secondary instabilities are
\begin{eqnarray}
\label{eq:euler_pi}
\partial_t\delta\bb{v} &+& \left(\Delta\bb{v}\bcdot\del\right)\delta\bb{v} +
\left(\delta\bb{v}\bcdot\del\right)\Delta\bb{v}  = 
- \del(\delta P + \Delta\bb{B} \bcdot \delta\bb{B})
\nonumber \\ 
&+& (\Delta\bb{B}\bcdot\del)\delta\bb{B} 
+ (\delta\bb{B}\bcdot\del)\Delta\bb{B} 
+ \nu \del^2{\delta \bb{v}} \,,
\\
\label{eq:induction_pi}
\partial_t \delta\bb{B} &+& 
\left(\Delta\bb{v} \bcdot \del \right)\delta\bb{B} +
\left(\delta \bb{v} \bcdot \del \right)\Delta\bb{B} =
\nonumber \\ 
&& \left(\Delta\bb{B} \bcdot \del \right) \delta\bb{v} +
\left(\delta\bb{B} \bcdot \del \right) \Delta \bb{v} + \eta \del^2{\delta\bb{B}} \,,
\end{eqnarray}
where $\Delta{\bb{v}} \equiv\bb{V}_0 \sin(Kz)$, $\Delta{\bb{B}} \equiv
\bb{B}_0 \cos(Kz)$, $\del \bcdot \delta\bb{v} = \del \bcdot
\delta\bb{B} =0$, and $\delta P$ stands for the pressure perturbation.
We seek solutions of the form
\begin{eqnarray}
\label{eq:v_sol}
\delta \bb{v}(\bb{x},t) &=& \delta{\bb{v}}_0(z) \, \exp{[s t - i\bb{k}\bcdot\bb{x}]} \,, \\ 
\label{eq:b_sol}
\delta \bb{B}(\bb{x},t) &=& \delta{\bb{B}}_0(z) \, \exp{[s t - i\bb{k}\bcdot\bb{x}]} \,,
\end{eqnarray}
where the amplitudes $\delta{\bb{v}}_0(z)$ and $\delta{\bb{B}}_0(z)$
are periodic in $z$ with period $2\pi/K$.  Substituting expressions
(\ref{eq:v_sol}) and (\ref{eq:b_sol}) into Equations
(\ref{eq:euler_pi}) and (\ref{eq:induction_pi}), and using the
divergenceless nature of the perturbed fields, we derive a set of
higher order differential equations for the vertical components of the
secondary velocity and magnetic fields,
\begin{eqnarray}
\label{eq:linear_sys_nu}
(s + \nu \mathcal{Q})\mathcal{Q}\delta v_z 
&-& i(\bb{k}_{\rm h}\bcdot\Delta{\bb{v}}) (\mathcal{Q}-K^2) \delta v_z \nonumber \\
&+& i(\bb{k}_{\rm h}\bcdot\Delta{\bb{B}}) (\mathcal{Q}-K^2) \delta B_z
=0 \,, \\
\label{eq:linear_sys_eta}
(s + \eta \mathcal{Q})\delta B_z 
&+& i(\bb{k}_{\rm h}\bcdot\Delta{\bb{B}}) \delta v_z 
- i(\bb{k}_{\rm h}\bcdot\Delta{\bb{v}}) \delta B_z = 0 \,, \,\,\,\,\,
\end{eqnarray}
where the horizontal wavevector $\bb{k}_{\rm h}$ is such that $\bb{k}
= \bb{k}_{\rm h} + k_z \hat{\bb{z}}$, and the differential operator
$\mathcal{Q} \equiv k_{\rm h}^2 -\partial_z^2$.  The ratio $k_z/K$
need not be rational but $0 \le k_z/K \le 1/2$ (GX94).  Equations
(\ref{eq:linear_sys_nu}) and (\ref{eq:linear_sys_eta}), with suitable
boundary conditions, pose an eigenvalue problem\footnote{The method of
  solution of these differential equations as well as a discussion of
  the physics of the secondary modes will be presented elsewhere.} for
the parasitic growth rate $s= s(\nu, \eta, K, B_0, k_z, \theta, k_{\rm
  h})$, where $\theta$ denotes the angle between the horizontal
wavevector $\bb{k}_h$ and the radial direction.

\citet[][hereafter LL07]{LL07} carried out a systematic study of the
saturation of the MRI in incompressible MHD with explicit viscosity
and resistivity.  They presented results for the dimensionless stress
at saturation for a grid of models with ${\rm Pm}=\{0.125, 0.25, 1.0,
4.0, 8.0\}$ and ${\rm Re} \equiv SL_z^2/\nu = \{200, 400, 800, 1600,
3200, 6400\}$, with $S=3\Omega_0/2$. To relate our results to those of
LL07, we translate their Reynolds number ${\rm Re}$ into $\Lambda_\nu
= 3{\rm Re}/2\beta$.  Their parameter $\beta \equiv
S^2L_z^2/\bar{v}_{{\rm A}z}^2$ is a proxy for the plasma $\beta$
parameter in a stratified disk with equivalent height $L_z$ (in our
units $L_z = 2\sqrt{\beta}/3$).  Setting $\beta=100$, as in LL07, it
follows that the values for ${\rm Re}$ cited above correspond to
$\Lambda_\nu = \{3, 6, 12, 24, 48, 96\}$.  Note that these are rather
small.

We solved equations (\ref{eq:linear_sys_nu}) and
(\ref{eq:linear_sys_eta}) for all the possible combinations of the set
of values of $\Lambda_\nu$ and ${\rm Pm}$ defined above and searched the
parameter space defined by $(B_0, k_z, \theta, k_{\rm h})$ in order to
identify the most relevant, fastest growing secondary modes.  We have
not been able to find unstable parasitic modes with $k_{\rm h} > K$.
This result generalizes the findings of GX94 to the non-ideal MHD
regime.

\section{Saturation of the Magnetorotational Instability}
\label{sec:saturation}

The secondary modes will be clearly dynamically important when their
growth rates are comparable to, or greater than, the growth rates of
the primary modes upon which they feed. We refer to this instance as
the  ``saturation'' of the primary MRI mode. It is then convenient to
define the saturation amplitude $B_0^{\rm sat}(\nu,\eta,K)$ as the
amplitude that the magnetic field produced by the MRI must have grown
to in order for the instantaneous growth rate of the fastest parasitic
mode, $s_{\rm max}(\nu,\eta,K)$, to match that of the primary,
$\Gamma(\nu,\eta,K)$\footnote{In neglecting the temporal dependence of
  the background, it is assumed that the secondary growth rates are
  large compared to the primary growth rate; thus our definition of
  saturation implies an extrapolation to the regime where this
  assumption is not strictly satisfied.}.

In the ideal limit, the growth rate of the secondary modes derived
from equations (\ref{eq:linear_sys_nu}) and (\ref{eq:linear_sys_eta})
is linear in the amplitude of the primary magnetic field (GX94). Thus,
the amplitude $B_0$ at which the growth rate of the fastest secondary
equals the growth rate of a given primary mode can be estimated after
solving these equations with $\nu=\eta=0$. However, in the non-ideal
case the amplitude $B_0$ cannot be scaled out of the problem, and the
growth rate of the secondary modes depends on it in a non-trivial
way. Figure \ref{fig:max_growths} shows the fastest growth rates
$s_{\rm max}(\nu,\eta, K_{\rm max}, k_z, \theta)$ of various secondary
modes that feed off the fastest primary MRI mode for three
combinations of $\Lambda_\nu$ and ${\rm Pm}$, with curves for several
values of $k_z$.  In all of the cases shown, $B_0 = B_0^{\rm
  sat}(\nu,\eta,K_{\rm max})$, i.e., the amplitude of the primary MRI
mode is such that the fastest secondary growth rate, $s_{\rm
  max}(\nu,\eta, K_{\rm max})=\Gamma_{\rm max}(\nu,\eta)$.

For all the cases within the explored dissipative regime, the fastest
parasitic modes are non-axisymmetric ($\theta \ne 0$), have the same
vertical periodicity as the primary mode ($k_z = 0$), and have purely
real growth rates. The fastest modes have horizontal
wavevectors that are nearly aligned with either the velocity or the magnetic
field of the primary ($\Delta{\bb{v}}$, $\Delta{\bb{B}}$).  
The first type are clearly related to
Kelvin-Helmholtz instabilities while the latter are related to tearing
modes. The ratio between the horizontal wavenumber of the fastest
parasitic mode, $k_{\rm h, max}(\nu,\eta,K_{\rm max})$, and the
wavenumber of the fastest MRI mode is rather insensitive to either
$\Lambda_\nu$ or ${\rm Pm}$; $k_{\rm h, max}/K_{\rm max}$ varies from
0.59, in ideal MHD, to 0.46, in the cases with highest viscosity and
resistivity.

We calculate the dimensionless stress at saturation as
$\alpha_{\rm sat} \equiv \bar{T}^{\rm sat}_{r\phi}/(SL_z)^2$, where
$\bar{T}^{\rm sat}_{r\phi}\equiv \bar{R}^{\rm
  sat}_{r\phi}-\bar{M}^{\rm sat}_{r\phi}$, is the sum of the Reynolds
and Maxwell stresses
\begin{eqnarray}
\bar{R}^{\rm sat}_{r\phi} &\equiv&\frac{1}{L_z}\int_{-L_z\!/2}^{L_z\!/2} 
V^{\rm sat}_{0,r}(z)V^{\rm sat}_{0,\phi}(z) dz \,, \\
\bar{M}^{\rm sat}_{r\phi} &\equiv&\frac{1}{L_z}\int_{-L_z\!/2}^{L_z\!/2} 
B^{\rm sat}_{0,r}(z)B^{\rm sat}_{0,\phi}(z) dz \,.
\end{eqnarray}
These expressions are integrated to obtain the dimensionless stress
$\alpha_{\rm sat}$ in terms of the parameter $\beta$,
\begin{eqnarray}
\alpha_{\rm sat} &=& \frac{1}{4\beta} 
[(V_0^{\rm sat})^2 \sin2\theta_{\rm V} -
( B_0^{\rm sat})^2 \sin2\theta_{\rm B}] \,.
\end{eqnarray}

The left panel of Figure \ref{fig:alpha_sat} shows $\alpha_{\rm sat}$
in the case where both primary and secondary instabilities evolve
unimpeded. There are competitive effects that set the value of
$\alpha_{\rm sat}$, which is dominated by the Maxwell stress.  As
dissipation increases, the saturation amplitude $B^{\rm sat}_0$
increases from $B^{\rm sat}_0=3.8$, in ideal MHD, to $B^{\rm
  sat}_0\simeq 5.5$, in the cases with high dissipation. However, the
angle $\theta_{\rm B}$ decreases toward $\pi/2$, so that
$|\sin2\theta_{\rm B}|$ decreases. Both effects roughly compensate
each other so that the final value of $\alpha_{\rm sat}$ changes only
by a factor of 2. Therefore, for the range of dissipation coefficients
that we explored, the saturation amplitude of the primary modes should
be insensitive to the dissipation coefficients if both the
fastest primary and secondary instabilities are permitted. The value
of $\alpha_{\rm sat}$ does not depend on ${\rm Pm}$ for large
$\Lambda_\nu$, and the range of ${\rm Pm}$ for which the results do
not depend on ${\rm Pm}$ seems to increase with $\Lambda_\nu$.

The right panel of Figure \ref{fig:alpha_sat} shows $\alpha_{\rm sat}$
when the primary and secondary modes considered are the fastest MRI
and parasitic modes that can fit in a domain $(L_r,L_{\phi},L_z)
= (1,4,1) \times 2\sqrt{\beta}/3$ with $\beta=100$, as considered in
LL07. The purpose of this exercise is to emulate the situation in
which a finite simulation domain might constrain the availability of
primary and/or secondary modes. The dependence of $\alpha_{\rm sat}$
on the dimensionless numbers $(\Lambda_\nu,{\rm Pm})$ is modified with
respect to the case where there are no limitations on the range of
primary or secondary modes; $\alpha_{\rm sat}$ decreases with
decreasing ${\rm Pm}$ for a wider range of $\Lambda_\nu$.  This trend
is similar to that observed by LL07, but the dependence on
${\rm Pm}$ is less pronounced and the predicted value of $\alpha_{\rm sat}$ is
smaller than that found in non-linear simulations (by a factor of
6 at ${\rm Pm} =1$).  Most of the differences in
$\alpha_{\rm sat}$ associated with
domain size are due to the limitations on the
primaries.  Although the fastest secondaries do not fit in the domain,
there are other secondary modes with comparable growth rates that can
lead to ``saturation'' at slightly larger primary amplitudes.

\section{Discussion}
\label{sec:discussion}

We have investigated the spectrum of parasitic instabilities that feed
off the MRI in viscous, resistive MHD, focusing our attention on the
parameter space currently accessible to numerical simulations. Our
study suggests that important differences between two-dimensional and
three-dimensional simulations are to be expected. The fastest
parasitic modes are non-axisymmetric and have the same vertical
periodicity as the primary upon which they feed. They tend to have
wavevectors that are almost aligned with the velocity or magnetic
fields generated by the MRI. As we will detail in a separate paper,
the fastest parasitic instabilities can be roughly grouped into
Kelvin-Helmholtz and tearing mode instabilities. The first type feed
off the sinusoidal velocity field of the MRI and are quenched when the
viscosity is increased. The second feed off the MRI currents and are
enabled by resistivity.

The values of $\alpha_{\rm sat}$ that result from the rather crude
procedure that we followed to find the amplitudes of the MRI fields at
saturation are similar, within factors of a few, to the values
obtained from numerical simulations in the turbulent regime. Our
results suggest, however, that the saturation amplitude of the MRI may
not be well determined if the simulation domain limits the available
primary and secondary modes. The domains should be large enough
vertically to accommodate the fastest primary mode, i.e., $L_z \ge
2\pi/K_{\rm max}(\nu,\eta)$, and should have aspect ratios that
allow the fastest parasitic modes, i.e., $L_r,L_\phi
\gtrsim 2 L_z$ (see \S \ref{sec:saturation}).

Having gained some insight into the dynamics of primary MRI modes and
their parasitic instabilities, it is instructive to review some recent
numerical results:

$\bullet$ LL07 carried out a series of shearing box simulations in
incompressible MHD with explicit dissipation.  For the range of
parameters that they explored, $0.1 \lesssim {\rm Pm} \lesssim 10$,
the stress at saturation decreases with decreasing magnetic Prandtl
number.  A weak dependence on the Reynolds number cannot be
discounted.  For several of the runs, the most unstable MRI mode does
not fit within the numerical domain, and neither do the fastest
parasitic modes.  It is worth asking whether the observed
trends of $\alpha_{\rm sat}$  with ${\rm Pm}$ and ${\rm Re}$ might
be biased by these constraints.

$\bullet$ \cite{MS08} explored viscous effects in two-dimensional
simulations. In several cases, ``saturation'' is not achieved since
the turbulent stresses are still increasing at the end of the
runs. Two different effects might be playing a role in the observed
behavior. Axisymmetric parasitic modes grow at only a fraction of the
rate of the fastest non-axisymmetric modes.  Kelvin-Helmholtz
parasites are further slowed by viscosity. Therefore, the MRI field
needs to grow to higher amplitudes before the secondary instabilities
can compete with the most unstable (available) primary mode.

$\bullet$ \cite{Bodoetal08} exposed a dependence of saturation on the
aspect ratio of the simulation domain. They found that the saturated
stresses decrease when $L_r/L_z$ varies from 1 to 4, and found less
significant differences between aspect ratios 4 and 8. We have shown
that even in non-ideal MHD, the fastest parasitic modes have
horizontal wavelengths roughly twice as large as the vertical
wavelength of the dominant primary MRI mode. It is tempting to
attribute Bodo et al's (2008) results to the exclusion of parasitic
modes at the $1:1$ aspect ratio.

While this Letter was in draft form, we learned of the investigations
of \citet[][hereafter LLB09]{LLB09}, which are similar to our own. A
brief comparison of methods and conclusions is in order. Both studies
generalize GX94 to resistive MHD and characterize the types and growth
rates of the non-ideal parasitic modes. LLB09 neglect viscous effects,
which we include, but supplement their analysis with direct numerical
simulations. A major conclusion of both studies is that parasitic
modes can be limited by the size of the simulation domain. Thus box
size may be as important as numerical resolution for the saturation of
the MRI.

LLB09 ultimately conclude, however, that parasitic modes are
unimportant for two reasons. First, they suggest that parasites should
``overtake'' the primaries only when the latter reach large non-linear
amplitudes, and then the gas behaves compressibly.  We have shown that
the secondary modes become dynamically important when the MRI magnetic
field is of the order of a few times the vertical field.  Therefore,
we argue that if the initial vertical field is sufficiently weak, then
the parasites can erupt before the magnetic pressure approaches the
gas pressure, although this regime is expensive to simulate with
compressible codes.  Second, LLB09 find that in large boxes---where
parasitic modes are possible---channel modes emerging from the
turbulent regime do not reach large amplitudes; they conclude that
something other than parasitic modes must be responsible for their
disruption. This may be true, although we suggest another explanation
below. While LLB09 speculate about multi-mode interactions, they offer
no definite or calculable alternative to parasitic modes. At present,
the properties of parasitic modes provide the only analytical
guidance, other than the linear MRI dispersion relation, to the choice
of dimensions for incompressible or weak-field simulations.

The numerical simulations seem to suggest that the primary modes reach
higher amplitudes than predicted in Section \ref{sec:saturation}.
Perhaps terms neglected from equations (\ref{eq:linear_sys_nu}) and
(\ref{eq:linear_sys_eta}) delay the onset of the secondary
instabilities by reducing their growth rates. Alternatively,
saturation may occur not when the secondaries achieve the same growth
rate as the primary, but when they achieve the same amplitude. The
latter occurs later when the MRI develops from a quiet start. Let
$t_g$ be the time at which the growth rates match, when the primary
has amplitude $B_0(t_g)$.  The instantaneous parasitic growth rate
$s_{\rm max} (t)\simeq [B_0(t)/B_0(t_g)]\Gamma_{\rm max} $ for $t\ge
t_g$. Suppose that the fastest parasite begins with amplitude
$\epsilon\ll1$ relative to the primary at $t_g$.  Then when secondary
and primary amplitudes match, at a later time $t_a$, the primary has
grown by a factor $B_0(t_a)/B_0(t_g)\simeq 1+\ln(1/\epsilon)$.  That
is, the equal-amplitude criterion predicts a saturation amplitude
larger by $\sim\ln(e/\epsilon)$ compared to our previous
equal-growth-rate criterion. LLB09 seeded their simulations with small
but unspecified noise, which set the initial amplitude of the
parasites. If $\epsilon \sim 10^{-4}$ then the overshoot factor
$\ln(e/\epsilon)\approx 10$. In the fully turbulent regime, the
parasites would start from a larger amplitude, say $\epsilon\sim 0.1$,
leading to a smaller overshoot $\ln(e/\epsilon)\sim 3$. Channel modes
are then less prominent in the turbulent state because of this smaller
peak amplitude as well as the larger background of other modes.
Further numerical investigations should shed light on these issues.

\acknowledgments{This Letter benefited from trenchant criticism by an
  anonymous referee.  M.E.P. is grateful to Chi-kwan Chan, Peter
  Goldreich, and Aldo Serenelli for useful discussions and gratefully
  acknowledges support from the Institute for Advanced Study.
  This work was supported in part by NSF award PHY-0821899 ``Center
  for Magnetic Self-Organization in Laboratory and Astrophysical
  Plasmas''.}

\end{document}